\documentclass[aps,pre,reprint,superscriptaddress,amsmath,showpacs,amssymb]{revtex4-2}
\usepackage{amssymb}
\usepackage{tipa}
\usepackage{bm,epsfig,mathrsfs,amsmath,amssymb,graphicx}
\usepackage{epsfig,amsmath,graphicx,amssymb}
\usepackage{graphicx}
\usepackage{dcolumn}
\usepackage{bm}
\usepackage{color}
\usepackage{tikz}
\usepackage{tikz-3dplot}
\usepackage{tkz-euclide}
\usetikzlibrary{decorations.markings}
\usepackage[toc,title,titletoc,header]{appendix}

\begin{document}
\newcommand\uleip{\affiliation{Institut f\"ur Theoretische Physik, Universit\"at Leipzig,  Postfach 100 920, D-04009 Leipzig, Germany}}
\newcommand\chau{\affiliation{ 
 Charles University,  
 Faculty of Mathematics and Physics, 
 Department of Macromolecular Physics, 
 V Hole{\v s}ovi{\v c}k{\' a}ch 2, 
 CZ-180~00~Praha, Czech Republic
}}

\title{Maximum efficiency of absorption refrigerators at arbitrary cooling power}
\author{Zhuolin Ye}\email{zhuolinye@foxmail.com}\uleip
\author{Viktor Holubec}\email{viktor.holubec@mff.cuni.cz}\uleip\chau

\begin{abstract}
We consider absorption refrigerators consisting of simultaneously operating Carnot-type heat engine and refrigerator. Their maximum efficiency at given power (MEGP) is given by the product of MEGPs for the internal engine and refrigerator. The only subtlety of the derivation lies in the fact that the maximum cooling power of the absorption refrigerator is not limited just by the maximum power of the internal refrigerator but, due to the first law, also by that of the internal engine. As a specific example, we consider the simultaneous absorption refrigerators composed of low-dissipation (LD) heat engines and refrigerators, for which the expressions for MEGPs are known. The derived expression for maximum efficiency implies bounds on the MEGP of LD absorption refrigerators. It also implies that a slight decrease in power of the absorption refrigerator from its maximum value results in a large nonlinear increase in efficiency observed in heat engines whenever the ratio of maximum powers of the internal engine and the refrigerator does not diverge. Otherwise, the increase in efficiency is linear as observed in LD refrigerators. Thus, in all practical situations, the efficiency of LD absorption refrigerators significantly increases when their cooling power is slightly decreased from its maximum. %We have tested all the analytical results obtained for the LD model by exact numerical optimization.
%\rightline{PACS number(s): 05.70.Ln, 05.20.−y, 07.20.Pe~~~~~~~~~~~~~~~~~~}
\end{abstract}

\maketitle

%%%%%%%%%%%%%%%%%%%%%%%%%%%%%%%%%%%%%%%%%%%%%%%%%%%%%%%%%%%%%%%%%%%%%%%%%%%%%%%%%%%%%%%%%%%%%%%%
\section{Introduction}
%%%%%%%%%%%%%%%%%%%%%%%%%%%%%%%%%%%%%%%%%%%%%%%%%%%%%%%%%%%%%%%%%%%%%%%%%%%%%%%%%%%%%%%%%%%%%%%%

The performance of heat engines, transforming heat to work, or refrigerators and heat pumps, displacing heat against a temperature gradient, is determined by two main quantities: output power and efficiency. Unfortunately, thermodynamic laws imply that they cannot be optimised simultaneously \cite{Callen2006}. This is because largest efficiencies correspond to reversible and thus slow processes, leading to output powers which are at best negligible fractions of the maximum power~\cite{PhysRevE.96.062107}. 

The implication for engineers, whose natural task is to develop designs that deliver a desired (fixed) power as cheap as possible, is that their devices in general do not operate in the regimes of maximum efficiency~\cite{Callen2006, Muller2007} or maximum power~\cite{Yvon1955, Chambadal1957, Novikov1958,PhysRevLett.105.150603, PhysRevLett.124.050603, PhysRevE.86.011127,PhysRevLett.95.190602, PhysRevLett.106.230602, PhysRevLett.112.180603,PhysRevE.91.052140, izumida2012efficiency,schmiedl2007efficiency, PhysRevLett.101.260601,PhysRevE.59.6448,Esposito2009,esposito2009thermoelectric,PhysRevE.81.041106,PhysRevB.82.235428}, which were both thoroughly investigated theoretically in the past, but rather in the regime with maximum efficiency corresponding to the given power (MEGP). The latter received the attention of the theory of finite-time thermodynamic processes only recently~\cite{Whitney2014,Whitney2015,PhysRevE.92.052125,holubec2016maximum, PhysRevE.93.050101,Dechant2017}, generalizing results obtained previously for a variety of trade-off relations between power and 
efficiency~\cite{apertet2013efficiency, yan1990class,PhysRevE.85.010104,PhysRevE.63.037102,angulo1991ecological,izumida2013coefficient, PhysRevE.87.012105, PhysRevE.82.051101,PhysRevE.93.032152,long2015ecological}.

Unlike model-independent equilibrium results such that maximum efficiency of thermal devices is the Carnot efficiency~\cite{Callen2006, Muller2007}, all available results on optimal performance of thermal devices operating with finite cycle times are based on specific model systems. Nevertheless, these models are usually constructed in an idealized fashion so that real-world devices inevitably dissipate more and thus operate at smaller efficiencies. 
The results for MEGP obtained in these models thus represent (loose) upper bounds on real-world efficiencies. 

Specifically, the idealized models just consider inevitable energy losses imposed by the second law of thermodynamics. In particular, losses connected to heat leakages and construction imperfections are neglected. Most of the idealized models operate along a finite-time Carnot cycle composed of two adiabatic and two isothermal branches and assume that the total entropy change in the universe during each of the isotherms obeys the so-called low-dissipation (LD) assumption~\cite{PhysRevLett.105.150603}
\begin{equation}
\Delta S_{\rm tot}= \Sigma/t,
\label{LD-diss}
\end{equation}
where the irreversibility parameter $\Sigma>0$ depends on details of the system construction, and $t$ is the duration of the isotherm. The low-dissipation assumption is not just a useful approximation allowing to derive explicit analytical results. This model exactly describes Brownian heat engines optimised with respect to output power~\cite{schmiedl2007efficiency,PhysRevE.92.052125}, which can now be realised in experiments~\cite{blickle2012realization,martinez2016brownian}. More generally, the low-dissipation model describes the first finite-time correction to the quasi-static dissipation, which was revealed not only in theoretical studies~\cite{sekimoto1997complementarity,PhysRevE.92.032113,cavina2017slow} but also in experiments \cite{martinez2016brownian, ma2019experimental}.
Furthermore, with respect to MEGP, the LD model was shown to be equivalent to the minimally nonlinear irreversible model~\cite{izumida2012efficiency, izumida2013coefficient, holubec2020maximum}, and, for small temperature gradients, also to the linear irreversible model~\cite{PhysRevE.93.050101}.
Regardless the relatively simple mathematical structure of the LD models, exact results on MEGP are so far known for LD heat engines~\cite{holubec2016maximum} and refrigerators~\cite{holubec2020maximum} only. Other devices such as absorption refrigerators~\cite{guo2019thermally} and heat pumps~\cite{guo2020equivalent} are still investigated numerically even when formulated within the LD setting. 

In the present paper, we consider absorption refrigerators consisting of simultaneously  operating Carnot-type (internal) heat engine and refrigerator. We show how the MEGP for this general model follows from the MEGPs for the internal heat engine and refrigerator. %Namely the Carnot-type LD absorption refrigerator composed of a LD refrigerator powered by a simultaneously operating Carnot-type LD heat engine.
To derive explicit results, we consider absorption refrigerators consisting of LD heat engines and refrigerators, for which expressions for MEGPs are known. The obtained MEGP represents a loose upper bound for efficiency of real-world absorption refrigerators, which recently experienced a renewed interest of physicists due to their potential to recycle waste heat in microscopic (quantum) devices~\cite{Levy2012,Correa2013,correa2014quantum,Brask2015,Gonzalez2017,Holubec2019}.

The rest of the paper is organized as follows. In Sec. \ref{germal-model-descri}, we introduce the general model and derive the general results. In Sec. \ref{sec-model}, we derive the MEGP for LD absorption refrigerators and discuss its properties. We conclude in Sec. \ref{sec-conclusion}.

% The expressions for maximum efficiencies of LD heat engines and refrigerators at given power are derived in Appendixes \ref{appx:opt-engine} and \ref{appx:opt-refrigerator}. Appendix \ref{appx:method-comparison}.

%%%%%%%%%%%%%%%%%%%%%%%%%%%%%%%%%%%%%%%%%%%%%%%%%%%%%%%%%%
\begin{figure}[tp]
\centering
\begin{tikzpicture}[scale=0.75]
%\draw[help lines] (-6,6) grid (6, 6); 

\begin{scope}[line width=0.5pt,decoration={
    markings,
    mark=at position 0.62 with {\arrow{>}}}
    ] 
    
    \draw[thick,black!60!green] (0.7,0) -- (5.1,0); %bath 
    \draw[thick,red] (0.7,4.4) -- (5.3,4.4);
    \draw[thick,blue] (0.7,-4.4) -- (5.3,-4.4);
    
    \draw(0.7,0) node[left,black!60!green] {$T_{\rm m}$}; %bath label 
    \draw(0.7,4.4) node[left,red] {$T_{\rm h}$};
    \draw(0.7,-4.4) node[left,blue] {$T_{\rm c}$};
    
    \draw[thick] (3,2.2) circle(1.2cm); %circle
    \draw[thick] (3,-2.2) circle(1.2cm);
    
    \draw(3,2.2) node[above] {$\text{Carnot}$}; %circle label
    \draw(3,1.55) node[above] {$\text{heat engine}$};
    \draw(3,-2.2) node[above] {$\text{Carnot}$};
    \draw(3,-2.85) node[above] {$\text{refrigerator}$};
    
    \draw[line width=1.5pt,red,->] (3,4.25) -- (3,3.55); % heat arrow
    \draw[line width=1.5pt,black!60!green,->] (3,0.85) -- (3,0.15);
    \draw[line width=1.5pt,blue,->] (3,-4.25) -- (3,-3.55);
    \draw[line width=1.5pt,black!60!green,->] (3,-0.85) -- (3,-0.15);
    
    \draw(3,3.9) node[right,red] {$N_{\rm e} Q_{\rm h}$}; % heat label
    \draw(3,0.5) node[right,black!60!green] {$N_{\rm e} Q_{\rm me}$};
    \draw(3,-3.9) node[right,blue] {$N_{\rm r} Q_{\rm c}$};
    \draw(3,-0.5) node[right,black!60!green] {$N_{\rm r} Q_{\rm mr}$};
    
    % \draw[thick] (5,2.2) -- (6.2,2.2); % work arrow
    % \draw[thick] (5,-2.2) -- (6.2,-2.2);
    % \draw[line width=1pt, black, postaction={decorate}](6.2,2.2)--(6.2,-2.2);
    
    % \draw[thick,postaction={decorate}] (4.2,2.3) arc (50:-50:3);% work circle
    
    \draw[line width=1.2pt,black!60!brown,->] (4.2,2.2) -- (5,2.2);
    \draw[line width=1.2pt,black!60!brown,->] (5,-2.2) -- (4.2,-2.2);
    \draw[dashed,line width=0.8pt,black!40!purple,-stealth] (5.35,2.1) arc (45:-45:3);% work circle
    
    \draw(5.35,0) node[right,black!40!purple] {$W$}; % work label
    \draw(4.1,2.5) node[right,black!60!brown] {$N_{\rm e} W_{\rm e}$};
    \draw(4.1,-1.9) node[right,black!60!brown] {$N_{\rm r} W_{\rm r}$};
    
\end{scope}
\end{tikzpicture}
\caption{Sketch of the Carnot absorption refrigerator (CAR) composed of internal Carnot heat engine and Carnot refrigerator. The overall CAR system communicates with three heat reservoirs at temperatures $T_{\rm h} > T_{\rm m}>T_{\rm c}$. Both the internal engine and refrigerator use as their heat sink the reservoir at the intermediate temperature $T_{\rm m}$. The engine in addition communicates with the hot bath at $T_{\rm h}$ and the refrigerator with the cold bath at $T_{\rm c}$.}
\label{fig:model}
\end{figure}
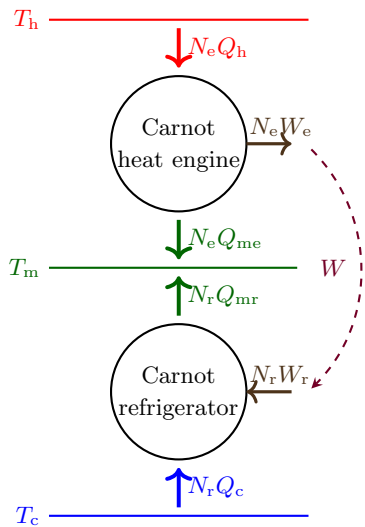
%%%%%%%%%%%%%%%%%%%%%%%%%%%%%%%%%%%%%%%%%%%%%%%%%%%%%%%%%%

%%%%%%%%%%%%%%%%%%%%%%%%%%%%%%%%%%%%%%%%%%%%%%%%%%%%%%%%%%%%%%%%%%%%%%%%%%%%%%%%%%%%%%%%%%%%%%%%
\section{Carnot absorption refrigerators}
\label{germal-model-descri}
%%%%%%%%%%%%%%%%%%%%%%%%%%%%%%%%%%%%%%%%%%%%%%%%%%%%%%%%%%%%%%%%%%%%%%%%%%%%%%%%%%%%%%%%%%%%%%%%

We consider absorption refrigerators consisting of a finite-time Carnot heat engine and refrigerator, which we call as Carnot absorption refrigerators (CARs). As shown in Fig.~\ref{fig:model}, the internal engine utilizes the temperature gradient $T_{\rm h} - T_{\rm m}>0$ between a hot \textcolor{black}{thermal} reservoir and a \textcolor{black}{thermal} reservoir at a medium temperature to generate work. This work is then used to propel the internal refrigerator, which pumps heat from the cold \textcolor{black}{thermal} reservoir at temperature $T_{\rm c}<T_{\rm m}$ into the intermediate bath. As a result, the CAR utilizes heat from the hot body to further cool the cold one. In practice, such refrigerators are often used in cases where there is no reliable source of electricity, for example in caravans. While described already in 1858 by Ferdinand Carr{\' e}, absorption refrigerators now acquired renewed attention in the field of quantum thermodynamics~\cite{Levy2012,Correa2013,correa2014quantum,Brask2015,Gonzalez2017,Holubec2019,Maslennikov2019}. This is because they seem to be promising building blocks of quantum devices, where they should help to keep the quantum parts at very low temperatures by utilizing the junk heat produced by classical chips inevitably present in these setups. 

In this work, we aim to provide an upper bound for MEGP for CARs and thus we assume that the internal engine and refrigerator work simultaneously~\footnote{One might argue that, when operating simultaneously, the internal heat engine cannot power the internal refrigerator because the engine generates work during half of the cycle only and accepts it during the rest. However, already realization of cyclically operating heat engine requires existence of a work source, where the work generated by the engine is stored and which is used to propel the engine during parts of the cycle where it accepts energy. The same work source can power the simultaneous CAR.
In practice, the energy transfer to/from the work source could lead to additional losses decreasing the performance of the CAR. Since our aim here is to provide an upper bound on efficiency, we neglect such losses. Furthermore, we are interested only on second law limitations on the performance of the CAR working under idealistic conditions (no heat leakages, no losses during work-to-work conversion) and thus the practical implementation of the work source is irrelevant for our analysis.}. Other possibility would be that they alternate~\cite{guo2019thermally}. Such CARs, however, involve during their operation idle periods of the internal devices and thus provide smaller MEGPs than the simultaneously operating setup. As we show below, the MEGP for simultaneous setups follows from MEGPs for the internal devices. In Appx.~\ref{appx:method-comparison}, we discuss that for the alternating setup the optimisation is actually more complicated and the knowledge of MEGPs of the internal engine and refrigerator is not sufficient for derivation of MEGP.

%%%%%%%%%%%%%%%%%%%%%%%%%%%%%%%%%%%%%%%%%%%%%%%%%%%%%%%%%%
\subsection{Working cycle of simultaneous CAR}
%%%%%%%%%%%%%%%%%%%%%%%%%%%%%%%%%%%%%%%%%%%%%%%%%%%%%%%%%%

Below, we will optimise the efficiency of the CAR with respect to durations $t_{\rm e}$ and $t_{\rm r}$ of the engine and refrigeration cycles and thus we assume that they are different. Duration of one cycle of the CAR, $t_{\rm s}$, is defined as a period after which both the internal devices attain their initial states. It is thus given by the least common multiple of $t_{\rm e}$ and $t_{\rm r}$. We assume that such a common multiple exists and denote as 
\begin{equation}
\textcolor{black}{N_{\rm e} = t_{\rm s}/t_{\rm e}}
\end{equation}
(\textcolor{black}{$N_{\rm r}=t_{\rm s}/t_{\rm r}$}) the number of engine (refrigeration) cycles performed per one full CAR cycle.

Now we are ready to define the thermodynamic quantities of interest, sketched in Fig.~\ref{fig:model}. Per CAR cycle, the engine produces work $W = N_{\rm e} W_{\rm e}$, which is used by the refrigerator to pump heat $N_{\rm r} Q_{\rm c}$ from the cold bath. Output power of the engine $W/t_{\rm s}$ and the input power of the refrigerator thus reads
\begin{equation}
P\equiv W_{\rm e}/t_{\rm e}= W_{\rm r}/t_{\rm r},
\label{first-law-of-thermodynamics}
\end{equation}
where $W_{\rm r} = W/N_{\rm r}$ denotes the work used by the refrigerator per refrigeration cycle. 

According to the first law, we have $W_{\rm e}= Q_{\rm h}-Q_{\rm me}$ and \textcolor{black}{$Q_{\rm c}=Q_{\rm mr} - W_{\rm r}$}. Here, $Q_{\rm h}$ and $Q_{\rm me}$ are the heats taken from the hot bath and delivered to the intermediate bath by the engine per period $t_{\rm e}$, respectively. Similarly, $Q_{\rm mr}$ is heat pumped into the intermediate bath by the refrigerator per period $t_{\rm r}$. The amount of heat extracted by the internal refrigerator from the cold bath per CAR cycle is given by $N_{\rm r} Q_{\rm c}$. The cooling power of the simultaneous CAR, $R_{\rm s}$, and the internal refrigerator, $R$, are thus the same and read
\begin{equation}
R_{\rm s} = R=N_{\rm r}Q_{\rm c}/t_{\rm s}=Q_{\rm c}/t_{\rm r}.
\label{power-tp}
\end{equation}
The energy input of the CAR is $N_{\rm e}Q_{\rm h}$ and thus its efficiency, referred to as the coefficient of performance (COP), is given by
\begin{equation}
\psi=\frac{N_{\rm r}Q_{\rm c}}{N_{\rm e}Q_{\rm h}}=\frac{Q_{\rm c}/t_{\rm r}}{Q_{\rm h}/t_{\rm e}}=\varepsilon\eta.
\label{efficiency}
\end{equation}
Here, $\eta=W_{\rm e}/Q_{\rm h}$ and $\varepsilon=Q_{\rm c}/W_{\rm r} = R/P$ denote the efficiency of the internal heat engine and refrigerator, respectively. 

%%%%%%%%%%%%%%%%%%%%%%%%%%%%%%%%%%%%%%%%%%%%%%%%
\subsection{Maximum cooling power}
\label{sec-general-results}
%%%%%%%%%%%%%%%%%%%%%%%%%%%%%%%%%%%%%%%%%%%%%%%%

Before we turn our attention to the MEGP for CARs, we determine the interval of allowed values of the cooling power~\eqref{power-tp}. Its minimum value 0 is achieved for infinitely slow cycles. The maximum cooling power, $R^*_{\rm s}$, turns out to be limited by maximum powers of both constituting devices.

\textcolor{black}{The power source of the refrigerator inside the CAR is the internal heat engine and thus the maximum cooling power of the CAR can not be larger than the maximum cooling power of the internal refrigerator without restrictions to input power, $R^*$, i.e. $R^*_{\rm s} \le R^*$.} Furthermore, the cooling power is related to output power of the engine by 
\begin{equation}
P= R/\varepsilon(R).
\label{second-relation}
\end{equation} 
\textcolor{black}{
We denote as $\bar{R}$ the maximum value of cooling power solving the equation
\begin{equation}
P^*=\bar{R}/\varepsilon(\bar{R}), \label{six-relation}
\end{equation}
where $P^*$ is the maximum power of the engine.} If $\bar{R} < R^*$, the engine is not powerful enough to utilize the whole potential of the refrigerator and \textcolor{black}{$R^*_{\rm s}=\bar{R}$}. Similarly, $\bar{R} > R^*$ means that the refrigerator is not powerful enough to use the whole power provided by the engine and \textcolor{black}{$R^*_{\rm s}=R^*$}. Altogether, we found that the maximum power of the CAR is given by
\begin{equation}
R_{\rm s}^*=\min (\bar{R}, R^*).
\label{determine-allowed-Rs}
\end{equation}
In the next section, we finally discuss the MEGP for the simultaneous CARs.

%%%%%%%%%%%%%%%%%%%%%%%%%%%%%%%%%%%%%%%%%%%%%%%%
\subsection{MEGP for simultaneous CARs}
\label{sec-general-maximum-COP}
%%%%%%%%%%%%%%%%%%%%%%%%%%%%%%%%%%%%%%%%%%%%%%%%

Inserting Eq.~\eqref{second-relation} for engine output power as function of power of the refrigerator in Eq.~\eqref{efficiency}, we obtain the COP of the CAR as function of $R$:
\begin{equation}
\psi(R)=\varepsilon(R)\eta\left[\frac{R}{\varepsilon(R)}\right].
\label{appxc-tilde-efficiency-present-model}
\end{equation}
To get the MEGP for CAR, we need to optimize the right-hand-side of this equation with respect to the durations of $t_{\rm r}$ and $t_{\rm e}$ for fixed $R$. 
\textcolor{black}{Using Eqs.~\eqref{second-relation}, \eqref{appa-power-engine}, and
\eqref{ini-appa-cooling-power-refrigerator}, Eq.~\eqref{appxc-tilde-efficiency-present-model}
can be rewritten in the form
\begin{equation}
\psi(R)=\frac{\eta_{\rm C}\varepsilon_{\rm C}}{1+\varepsilon_{\rm C}T_{\rm m}\sigma/R},  
\end{equation}
where $\eta_{\rm C}$ and $\varepsilon_{\rm C}$ are Carnot efficiency of reversible Carnot heat engine and refrigerator,
respectively, i.e. $\eta_{\rm C}=1-T_{\rm m}/T_{\rm h}$, and $\varepsilon_{\rm C}=T_{\rm c}/(T_{\rm m}-T_{\rm c})$, and \begin{equation}
\sigma = \Delta S_{\rm tot, r}/t_{\rm r} + \Delta S_{\rm tot, e}/t_{\rm e}
\label{eq:sigma}
\end{equation}
is the sum of average entropy production rates in the internal heat engine and internal refrigerator and thus the total average entropy production rate during the CAR cycle. Expressions for the total entropy changes per engine and refrigeration cycle, $\Delta S_{\rm tot, r}$ and $\Delta S_{\rm tot, e}$ are given in Eqs.~\eqref{deltaS-he} and \eqref{deltaS-re} in the appendix. Maximization of COP~\eqref{appxc-tilde-efficiency-present-model} at fixed $R$ is thus equivalent to minimization of the average entropy production rate~$\sigma = \sigma(R)$ under the same conditions.}

The internal heat engine depends on the setup and performance of the refrigerator through the refrigeration power $R/\varepsilon(R)$ only. Thus, in order to yield the maximum value of the product in Eq. \eqref{appxc-tilde-efficiency-present-model}, $\eta(R/\varepsilon(R))$ must attain its maximal value, $\eta^{\rm opt}(R/\varepsilon(R))$, corresponding to the given refrigeration power (MEGP). Furthermore, all known expressions for MEGP are decreasing functions of power~\cite{holubec2016maximum, holubec2020maximum,PhysRevE.93.050101,long2018performance,PhysRevE.94.052114}. Importantly, all these models 
neglect losses, which cannot be avoided by quasi-static operation, such as heat leakages, and thus they can saturate the Carnot bound on efficiency in the limit of vanishing power. Assuming that this idealization holds also in our present case, $\eta^{\rm opt}(R/\varepsilon(R))$ will be maximal if $\varepsilon(R)$ will be given by the maximum refrigerating efficiency at the given power, $\varepsilon^{\rm opt}(R)$. Altogether, the MEGP for the considered idealized CARs reads
\begin{equation}
\psi^{\rm opt}(R)=\varepsilon^{\rm opt}(R)\eta^{\rm opt}\left[\frac{R}{\varepsilon^{\rm opt}(R)}\right].
\label{opt-opt-relation-engine-refrigerator}
\end{equation}

The MEGP for the simultaneous CAR thus in general follows from the expressions for MEGPs for the internal engine and refrigerator. Let us now consider the simultaneous CAR composed of a LD heat engine and LD refrigerator \cite{holubec2016maximum, holubec2020maximum}. \textcolor{black}{For this specific model, we verified validity of Eq.~\eqref{opt-opt-relation-engine-refrigerator} by direct numerical maximisation of Eq.~\eqref{appxc-tilde-efficiency-present-model}. In the next section, we utilise known analytical expressions for $\eta^{\rm opt}$ and $\varepsilon^{\rm opt}$ for this model to discuss in detail properties of the MEGP~\eqref{appxc-tilde-efficiency-present-model} for this LD CAR based on analytical grounds.}

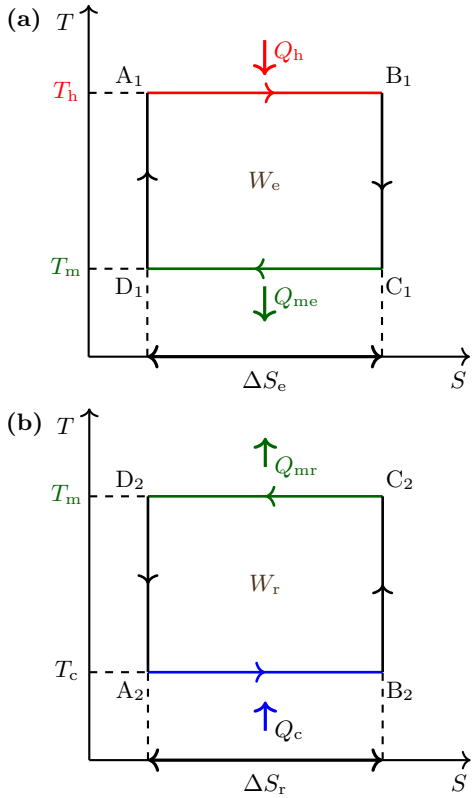
\begin{figure}[tp]
\centering
\begin{tikzpicture}[scale=0.78]

\begin{scope}[thick,decoration={
    markings,
    mark=at position 0.5 with {\arrow{<}}}
    ] 
    
    \draw[thick,->] (0,0) -- (0,6); %axis
    \draw[thick,->] (0,0) -- (6.5,0);
    
    \draw[very thick,->, black!60!green] (3,1.2) -- (3,0.6);
    \draw[very thick,->, red] (3,5.4) -- (3,4.8);
    \draw(3,1) node[right] {\textcolor{black!60!green}{$Q_{{\rm me}}$}};%axis label
    \draw(3,5.2) node[right] {\textcolor{red}{$Q_{\rm h}$}}; 
    \draw(3,3) node {\textcolor{black!60!brown}{$W_{\rm e}$}}; 

    \draw[dashed] (0,4.5) -- (1,4.5); %dashed line
    \draw[dashed] (0,1.5) -- (1,1.5);
    \draw[dashed] (1,0) -- (1,1.5);
    \draw[dashed] (5,0) -- (5,1.5);
  
    \draw[line width=1pt, black!60!green, postaction={decorate}](1,1.5)--(5,1.5);%cycle
    \draw[line width=1pt, red, postaction={decorate}](5,4.5)--(1,4.5);
    \draw[line width=1pt, black, postaction={decorate}](1,4.5)--(1,1.5);
    \draw[line width=1pt, black, postaction={decorate}](5,1.5)--(5,4.5);

    \draw(0.7,1.5) node[below] {$\rm D_1$};%cycle label 
    \draw(5.3,1.5) node[below] {$\rm C_1$}; 
    \draw(5.3,5.1) node[below] {$\rm B_1$}; 
    \draw(0.7,5.1) node[below] {$\rm A_1$};

    \draw(-0.38,4.85) node[below] {\textcolor{red}{$T_{\rm h}$}};%axis label
    \draw(-0.38,1.85) node[below] {\textcolor{black!60!green}{$T_{\rm m}$}}; 
    \draw(6.3,-0.1) node[below] {$S$}; 
    \draw(-0.4,6) node[below] {$T$}; 
    \draw[very thick,<->] (1,0) -- (5,0);
    \draw(3,-0.1) node[below] {$\Delta S_{\rm e}$};
    
    \draw(-1.1,6.1) node[below] {\textbf{(a)}}; 
\end{scope}

\end{tikzpicture}

\begin{tikzpicture}[scale=0.78]
\begin{scope}[thick,decoration={
    markings,
    mark=at position 0.5 with {\arrow{>}}}
    ] 
    
    \draw[thick,->] (0,0) -- (0,6); %axis
    \draw[thick,->] (0,0) -- (6.5,0);
    
    \draw[very thick,->, blue] (3,0.5) -- (3,1.0);
    \draw[very thick,->, black!60!green] (3,5.0) -- (3,5.5);
    \draw(3,0.5) node[right] {\textcolor{black}{$Q_{\rm c}$}};%axis label
    \draw(3,5) node[right] {\textcolor{black!60!green}{$Q_{{\rm mr}}$}}; 
    \draw(3,3) node {\textcolor{black!60!brown}{$W_{\rm r}$}}; 

    \draw[dashed] (0,4.5) -- (1,4.5); %dashed line
    \draw[dashed] (0,1.5) -- (1,1.5);
    \draw[dashed] (1,0) -- (1,1.5);
    \draw[dashed] (5,0) -- (5,1.5);
  
    \draw[line width=1pt, blue, postaction={decorate}](1,1.5)--(5,1.5);%cycle
    \draw[line width=1pt, black!60!green, postaction={decorate}](5,4.5)--(1,4.5);
    \draw[line width=1pt, black, postaction={decorate}](1,4.5)--(1,1.5);
    \draw[line width=1pt, black, postaction={decorate}](5,1.5)--(5,4.5);

    \draw(0.7,1.5) node[below] {$\rm A_2$};%cycle label 
    \draw(5.3,1.5) node[below] {$\rm B_2$}; 
    \draw(5.3,5.1) node[below] {$\rm C_2$}; 
    \draw(0.7,5.1) node[below] {$\rm D_2$};

    \draw(-0.38,4.85) node[below] {\textcolor{black!60!green}{$T_{\rm m}$}};%axis label
    \draw(-0.38,1.85) node[below] {\textcolor{black}{$T_{\rm c}$}}; 
    \draw(6.3,-0.1) node[below] {$S$}; 
    \draw(-0.4,6) node[below] {$T$}; 
    \draw[very thick,<->] (1,0) -- (5,0);
    \draw(3,-0.1) node[below] {$\Delta S_{\rm r}$};
    
    \draw(-1.1,6.1) node[below] {\textbf{(b)}}; 
\end{scope}
\end{tikzpicture}

\caption{Bath temperature-system entropy ($T$-$S$) diagrams of the components of the CAR depicted in Fig.~\ref{fig:model} considered in its low-dissipation version. (a) LD Carnot heat engine and (b) LD Carnot refrigerator. The horizontal colored lines are isotherms and the vertical black lines represent adiabats. The areas enclosed of the two rectangles equal to the respective works only if the cycles are realized quasi-statically.}
\label{fig:T-S}
\end{figure}

%%%%%%%%%%%%%%%%%%%%%%%%%%%%%%%%%%%%%%%%%%%%%%%%
%%%%%%%%%%%%%%%%%%%%%%%%%%%%%%%%%%%%%%%%%%%%%%%%
\section{Low-dissipation simultaneous CARs}
\label{sec-model}
%%%%%%%%%%%%%%%%%%%%%%%%%%%%%%%%%%%%%%%%%%%%%%%%
%%%%%%%%%%%%%%%%%%%%%%%%%%%%%%%%%%%%%%%%%%%%%%%%

Let us now consider the Carnot LD heat engine and refrigerator depicted in Fig.~\ref{fig:T-S}, for which the MEGPs were derived in Refs.~\cite{holubec2016maximum} and \cite{holubec2020maximum}, respectively. Their working cycles are composed of two isotherms realized in finite time and described by the irreversibility parameters $\Sigma_{\rm i}$, $i = $ h, me, c, mr. These isotherms are interconnected by infinitely fast adiabats~\footnote{\textcolor{black}{Adiabatic processes are described by vanishing average heat flux into the system and can be realized either by disconnecting the system from the thermal environment or by changing the control parameters in such a way that the average heat flux vanishes~\cite{Roldan2015,martinez2016brownian}. A fast adiabat is effectively an example of the former since the heat does not have enough time to enter the system, see Ref.~\cite{blickle2012realization} for an experimental realization of such branches in an over-damped Brownian setup. In general, fast changes of control parameters can bring the system far from equilibrium. In the overdamped description, this can be avoided by suitably tuning the jump in potential and temperature~ \cite{schmiedl2007efficiency,PhysRevE.92.052125}. However, in the underdamped regime this is in general not possible and one has to employ adiabatic branches with finite duration. In our present analysis, this would lead to a decrease in output power of all the considered devices.}}.

The internal engine accepts heat \begin{equation}
 Q_{\rm h}=T_{\rm h}\Delta S_{\rm e}-\frac{\Sigma _{\rm h}}{t_{\rm h}}
\label{qh}
\end{equation} 
during the hot isotherm (red) of duration $t_{\rm h}$ and \textcolor{black}{releases} heat 
\begin{equation}
 Q_{\rm me}=T_{\rm m}\Delta S_{\rm e}+\frac{\Sigma _{\rm me}}{t_{\rm me}}
 \label{qoh}
\end{equation} 
during the isotherm corresponding to the medium temperature (green) of duration $t_{\rm me}$. The terms proportional to the increase in the entropy of the working medium of the engine during the hot isotherm, $\Delta S_{\rm e}$, correspond to the reversible parts of the transferred heats. The total duration of the engine working cycle reads $t_{\rm e}= t_{\rm h}+t_{\rm me}$. Similarly, the refrigerator accepts heat 
\begin{equation}
Q_{\rm c}=T_{\rm c}\Delta S_{\rm r}-\frac{\Sigma_{\rm c}}{t_{\rm c}}
\label{qc}
\end{equation}
during the cold isotherm (blue) of duration $t_{\rm c}$ and \textcolor{black}{dumps} heat \begin{equation}
Q_{\rm mr}=T_{\rm m}\Delta S_{\rm r}+\frac{\Sigma_{\rm mr}}{t_{\rm mr}}
\label{qoc}
\end{equation}
during the intermediate isotherm (green) of duration $t_{\rm mr}$. The reversible components of transferred heats are proportional to the increase in the entropy of the working medium of the refrigerator during the cold isotherm, $\Delta S_{\rm r}$, which can be different than $\Delta S_{\rm e}$. The total duration of the refrigeration cycle is $t_{\rm r}= t_{\rm c}+t_{\rm mr}$. The internal heat engine and refrigerator operate reversibly if duration of all the isotherms diverge or if all the irreversibility parameters vanish.

Let us now consider a simultaneous CAR composed of the LD heat engine and LD refrigerator. We call it as LD simultaneous CAR. In what follows, we discuss in detail its performance in terms of MEGP.

%%%%%%%%%%%%%%%%%%%%%%%%%%%%%%%%%%%%%%%%%%%%%%%%
\subsection{MEGP}
%%%%%%%%%%%%%%%%%%%%%%%%%%%%%%%%%%%%%%%%%%%%%%%%

The MEGP for the LD CAR follows from Eq.~\eqref{opt-opt-relation-engine-refrigerator} after inserting the expressions for MEGP of the internal LD heat engine, $\eta^{\rm opt}$, and refrigerator, $\varepsilon^{\rm opt}$. For the engine, we derive $\eta^{\rm opt}$ in the Appx.~\ref{appx:opt-engine}. Similarly as the derivation given in Ref.~\cite{holubec2016maximum}, our present approach involves an approximation in calculation of the optimal redistribution of the total cycle duration between the two isothermal branches. Nevertheless, our analytical result for $\eta^{\rm opt}$ is, within the numerical precision, indistinguishable from the corresponding result obtained by exact numerical optimisation of the efficiency. For the refrigerator, we review in Appx.~\ref{appx:opt-refrigerator} the derivation of analytical expression for $\varepsilon^{\rm opt}$ from Ref.~\cite{holubec2020maximum}.

All results for MEGP available in the literature~\cite{long2018performance, PhysRevE.92.052125, PhysRevE.94.052114, PhysRevE.93.050101, holubec2016maximum, holubec2020maximum,PhysRevE.96.062107} are given as functions of the dimensionless variable
\begin{equation}
\delta X=\frac{X-X^*}{X^*},
\label{coordinate}
\end{equation}
measuring how much power is lost by operating the device at 
power $X$ smaller than the maximum power $X^*$. In our case, we have three such variables: the loss in power of the internal engine, $\delta P$, the loss in cooling power of the internal refrigerator, $\delta R$, and the loss in cooling power of the CAR, $\delta R_{\rm s}$. In general, these variables can assume values from the interval $[-1,0]$. The minimum is attained if the actual power is negligible compared to the maximum power and the maximum corresponds to devices operating at maximum power. However, in our specific setting where the input power of the refrigerator can be limited by the output power of the engine, the upper bound for $\delta R$ reads $R_{\rm s}^*/R^*-1 \le 0$.

In order to insert the known results for MEGP of the refrigerator and heat engine into Eq.~\eqref{opt-opt-relation-engine-refrigerator}, we need to express them in terms of refrigeration power $R$ and engine output power $P = R/\varepsilon^{\rm opt}(R)$, respectively. From now on, we use the shorthand notation $\varepsilon^{\rm opt}(\delta R) \equiv \varepsilon^{\rm opt}[R(\delta R)]$, where $R(\delta R) = (1 + \delta R) R^*$, and similarly for $\eta^{\rm opt}(\delta P)$.
Furthermore, in order to be able to discuss the MEGP of the CAR, $\psi^{\rm opt}=\varepsilon^{\rm opt}(\delta R)\eta^{\rm opt}(\delta P)$, as a function of the loss in cooling power of the CAR, we use Eqs. \eqref{second-relation} and \eqref{coordinate} to express $\delta P$ and $\delta R$ in terms of $\delta R_{\rm s}$. The result is
\begin{eqnarray}
\delta P&=&\frac{1}{\tilde{P}^*}\frac{1+\delta R}{\varepsilon^{\rm opt}(\delta R)}-1,
\label{third-relation}\\
\delta R&=&(1+\delta R_{\rm s})\tilde{R}_{\rm s}^*-1,
\label{delta-q-delta-r}
\end{eqnarray}
where we introduced the reduced maximum powers of the engine, $\tilde{P}^* = P^*/R^*$, and the CAR, $\tilde{R}_{\rm s}^* = R_{\rm s}^*/R^*$, measured in units of maximum power of the internal refrigerator.

When expressed in terms of $\delta P$, the MEGP of the LD heat engine, $\eta^{\rm opt}$, depends only on the ratio of the irreversibility parameters, $\Sigma_{\rm e}=\Sigma_{\rm h}/\Sigma_{\rm me}$, Carnot efficiency, $\eta_{\rm C}$, and $\delta P$. For details, see Appx.~\ref{appx:opt-engine}. Similarly, we show in Appx.~\ref{appx:opt-refrigerator} that $\varepsilon^{\rm opt}$ is only a function of $\Sigma_{\rm r}=\Sigma_{\rm mr}/\Sigma_{\rm c}$, $\varepsilon_{\rm C}=T_{\rm c}/(T_{\rm m}-T_{\rm c})$, and $\delta R$. \textcolor{black}{Since the MEGP, $\varepsilon^{\rm opt}(R)$, is a monotonously decreasing function of $R$, the ratio $R/\varepsilon^{\rm opt}(R)$ attains its maximum value for $R^*$. Therefore, Eqs.~\eqref{six-relation} and
\eqref{determine-allowed-Rs} imply that the reduced maximum power of the CAR $\tilde{R}_{\rm s}^*$ is given by 
\begin{equation}
\tilde{P}^*=\tilde{R}^*_{\rm s}/\varepsilon^{\rm opt}(\tilde{R}^*_{\rm s}-1),
\label{rsx-rx}
\end{equation}
if the resulting $\tilde{R}^*_{\rm s}$ is smaller than one, and by $\tilde{R}^*_{\rm s} = 1$ otherwise.} Hence $\tilde{R}^*_{\rm s}$ is determined by $\Sigma_{\rm r}$, $\varepsilon_{\rm C}$, and $\tilde{P}^*$. Collecting all these results and inserting them into Eq.~\eqref{opt-opt-relation-engine-refrigerator}, we can finally
write the MEGP of the LD simultaneous CAR 
in terms of the relative loss in its maximum cooling power, $\delta R_{\rm s}$. The resulting expression depends on the six parameters introduced above, namely
\begin{equation}
\psi^{\rm opt}\equiv\psi^{\rm opt}(\delta R_{\rm s}, \tilde{P}^*, \Sigma_{\rm e}, \Sigma_{\rm r}, \eta_{\rm C}, \varepsilon_{\rm C}).
\label{final-COP-abso-refri}
\end{equation}
In the following sections, we use this expression to provide more explicit results on MEGP of CARs.

%%%%%%%%%%%%%%%%%%%%%%%%%%%%%%%%%%%%%%%%%%%%%%%%
\subsection{Bounds on MEGP}
\label{section-4a}
%%%%%%%%%%%%%%%%%%%%%%%%%%%%%%%%%%%%%%%%%%%%%%%%

We start by deriving maximum and minimum values of the optimal COP $\psi^{\rm opt}$ with respect to working medium of the CAR (or of its constituents). In the LD approximation, the detailed physics of the working medium is described by the irreversibility parameters $\Sigma_{\rm i}$, $i = $ h, me, c, mr defined by Eqs.~\eqref{qh}--\eqref{qoc}~\cite{schmiedl2007efficiency,sekimoto1997complementarity,PhysRevE.92.052125,PhysRevLett.124.110606,abiuso2020geometric}. 

The optimal COP~\eqref{final-COP-abso-refri} depends on irreversibility parameters through the ratios $\Sigma_{\rm e}=\Sigma_{\rm h}/\Sigma_{\rm me}$ and $\Sigma_{\rm r}=\Sigma_{\rm mr}/\Sigma_{\rm c}$ and the reduced maximum power of the internal engine $\tilde{P}^*$. With respect to the former two, the optimal COP attains its minimum for $\Sigma_{\rm e}\to 0$ (hot isotherm of the internal engine cycle is reversible compared to the other one) and $\Sigma_{\rm r}\to\infty$ (cold isotherm of the refrigeration cycle is reversible compared to the other one). Its maximum $\psi^{\rm opt}$ is attained in the opposite limit $\Sigma_{\rm e}\to\infty$ and $\Sigma_{\rm r}\to 0$.
Taking these limits into Eq.~\eqref{final-COP-abso-refri}, we find the lower and upper bounds for the optimal COP:
\begin{equation}
0\le\psi^{\rm opt}\le\frac{\varepsilon_{\rm C}(1+\sqrt{-\delta R})}{2+\varepsilon_{\rm C}(1-\sqrt{-\delta R})}\frac{\eta_{\rm C}(1+\sqrt{-\delta P})}{2-\eta_{\rm C}(1-\sqrt{-\delta P})}.
\label{ultimate-bounds}
\end{equation}
This inequality has to be further optimized with respect to the parameter $\tilde{P}^*$, which enters the upper bound through Eqs.~\eqref{third-relation} and \eqref{delta-q-delta-r} for $\delta P$ and $\delta R$, respectively. 
Note that due to the limits $\Sigma_{\rm e}\to\infty$ and $\Sigma_{\rm r}\to 0$ taken to derive the upper bound, we have to use $\varepsilon^{\rm opt}=\varepsilon_+^{\rm opt}$ defined in Eq. \eqref{eta-duration2} in the formula for $\delta P$.
%The regime how theses limiting values of $\Sigma_{\rm e}$ and $\Sigma_{\rm r}$, leading to the bounds \eqref{ultimate-bounds}, can be (physically) realized is referred to Ref. \cite{holubec2020maximum}.}
One finds that the upper bound is a monotonously decreasing function of $\tilde{P}^*$ and thus its maximum is obtained for $\tilde{P}^*=0$. The resulting ultimate bounds on the optimal COP of the CAR at given cooling power read
\begin{equation}
0\le\psi^{\rm opt}\le\frac{\varepsilon_{\rm C}\eta_{\rm C}(1+\sqrt{-\delta R_{\rm s}})}{2-\eta_{\rm C}(1-\sqrt{-\delta R_{\rm s}})}\equiv\psi_+^{\rm opt}(\delta R_{\rm s}).
\label{extremely-ultimate-bounds}
\end{equation}
The upper bound evaluated for $\delta R_{\rm s}=0$, $\psi^{\rm opt}_+(0)=\varepsilon_{\rm C} \eta_{\rm C}/(2-\eta_{\rm C})$, denotes the upper bound for COP of the CAR at maximum cooling power. 

The increase in COP gained after a slight decrease of the cooling power from its maximum value can be measured by the expression
\begin{equation}
\frac{\psi_+^{\rm opt} (\delta R_{\rm s})-\psi_+^{\rm opt}(0)}{\psi_+^{\rm opt}(0)}=\frac{2-2\eta_{\rm C}}{2-\eta_{\rm C}}\sqrt{-\delta R_{\rm s}}+\mathcal{O}(\delta R_{\rm s}).
\label{relative-gain}
\end{equation}
Its derivative with respect to $\delta R_{\rm s}$ diverges, implying that a slight
decrease of the cooling power leads to a significant gain in the upper bound on COP. Qualitatively the same behavior has generally been observed for MEGPs of various heat engines \cite{PhysRevE.94.052114, PhysRevE.93.050101, holubec2016maximum, Dechant2017, Whitney2014,Whitney2015}. With respect to LD refrigerators,
the MEGP is proportional to $\sqrt{-\delta R}$ for a limited range of parameters only and behaves as $\propto -\delta R$ otherwise \cite{holubec2020maximum}. In the next section, we investigate whether the increase of MEGP for the CAR behaves for small values of $\delta R_{\rm s}$ always like the MEGPs in heat engines \cite{holubec2016maximum} or if it sometimes also exhibits the linear behavior observed in refrigerators~\cite{holubec2020maximum}.

%%%%%%%%%%%%%%%%%%%%%%%%%%%%%%%%%%%%%%%%%%%%%%%%
\subsection{MEGP near maximum cooling power}
%%%%%%%%%%%%%%%%%%%%%%%%%%%%%%%%%%%%%%%%%%%%%%%%

Examples of parameter regimes where the MEGP for LD refrigerators exhibits the two qualitatively different behaviors are $\Sigma_{\rm r}\to 0$ (square root) and $\Sigma_{\rm r}\to\infty$ (linear) \cite{holubec2020maximum}. We thus investigate behavior of the MEGP for the CAR~\eqref{final-COP-abso-refri} in these two regimes \textcolor{black}{using the cumbersome analytical expressions derived in Appendixes \ref{appx:opt-engine} and \ref{appx:opt-refrigerator}}.

\subsubsection{$\Sigma_{\rm r}\to 0$}
Expanding the exact expression for $\varepsilon^{\rm opt}$ in Eq. \eqref{eta-opt-appb-m} up to the first order with respect to $\Sigma_{\rm r}$ , we obtain
\begin{equation}
\varepsilon^{\rm opt}=\varepsilon^{\rm opt}_+-\frac{2(1+\varepsilon_{\rm C})\left(\varepsilon_+^{\rm opt}\right)^2(1-\sqrt{-\delta R})\sqrt{\tilde \Sigma_{\rm r}}}{\varepsilon_{\rm C}(-\delta R)^{1/4}(1+\sqrt{-\delta R})},
\label{epsilon-small}
\end{equation}
where $\tilde \Sigma_{\rm r}$ is defined below Eq. \eqref{max-cop-derive} and $\varepsilon^{\rm opt}_+$ in Eq. \eqref{eta-duration2}.
Substituting Eqs. \eqref{epsilon-small} and \eqref{appa-eta-opt-del-power-f3} for $\varepsilon^{\rm opt}$ and $\eta^{\rm opt}$ into Eq. \eqref{opt-opt-relation-engine-refrigerator} for MEGP for the CAR, expressing $\delta P$ and $\delta R$ in terms of $\delta R_{\rm s}$ using Eqs.~\eqref{third-relation} and \eqref{delta-q-delta-r}, and expanding the resulting expression up to the first order in $\delta R_{\rm s}$, we find
\begin{equation}
\psi^{\rm opt}=r_1 +r_2\sqrt{-\delta R_{\rm s}}.
\label{small-sigma-re-psi}
\end{equation}
The coefficients $r_1$ and $r_2$ depend in a complicated way on the parameters $\tilde{P}^*$, $\varepsilon_{\rm C}$, $\eta_{\rm C}$, and $\Sigma_{\rm e}$. The obtained dependence of the MEGP of the CAR on the loss in cooling power %  behavior~\eqref{small-sigma-re-psi} holds for arbitrary values of $\tilde{P}^*$, $\varepsilon_{\rm C}$, $\eta_{\rm C}$, and $\Sigma_{\rm e}$. , and it 
might have been expected since, in this parameter regime, the behavior near maximum power of the engine and the refrigerator is the same~\cite{holubec2016maximum,holubec2020maximum}.

\begin{figure}[tp]
\centering
\includegraphics[trim={0cm 0.4cm 0cm 0}, width=0.98\columnwidth]{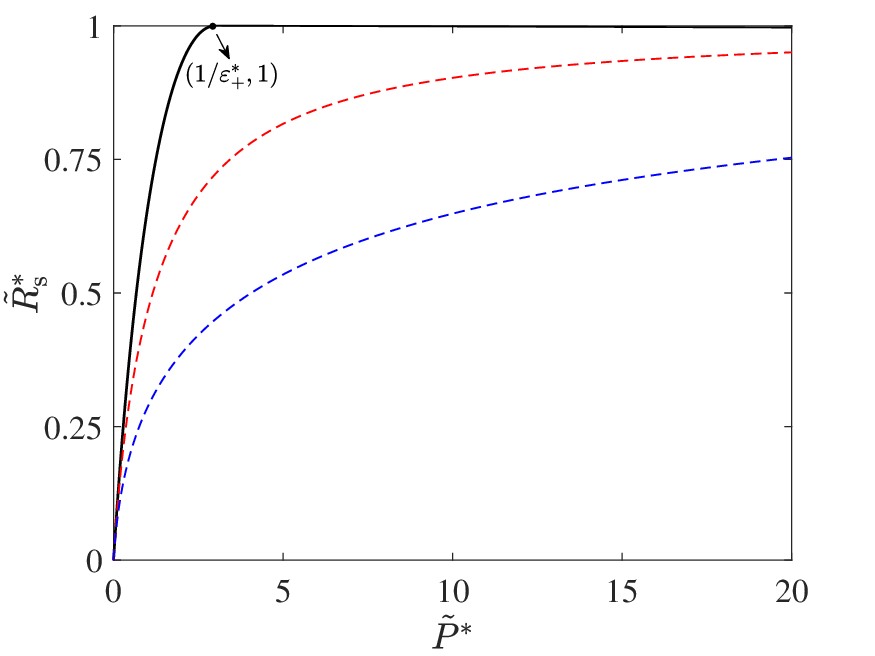}
\caption{The reduced maximum power of the CAR $\tilde{R}_{\rm s}^*$ as a function of the reduced maximum power of the internal heat engine $\tilde{P}^*$ for three values 0, 1, and 10 of the ratio $\Sigma_{\rm r}$ of irreversibility parameters, which increases from the uppermost solid
line to the lowermost dashed one. %The figure shows that $\tilde{R}_{\rm s}^*=1$ occurs for $\tilde{P}^*=1/\varepsilon_+^*$ when $\Sigma_{\rm r}=0$ (solid line) and for $\tilde{P}^*\to\infty$ when $\Sigma_{\rm r}>0$ (dashed lines). 
We take $\varepsilon_{\rm C}=1$.}
\label{fig:lambda}
\end{figure}

\subsubsection{$\Sigma_{\rm r}\to\infty$}
In this limit, the MEGP for LD refrigerators \eqref{eta-opt-appb-m} reads [see Eq. (29) in Ref. \cite{holubec2020maximum}]
\begin{equation}
\varepsilon^{\rm opt}\approx\frac{\delta R(1-\delta R)\varepsilon_{\rm C}}{2\delta R+(1+\delta R)(\delta R-\Sigma_{\rm r})\varepsilon_{\rm C}}.
\label{large-sigma-re}
\end{equation}
Using a similar procedure as for obtaining Eq.~\eqref{small-sigma-re-psi}, we find that up to the second order in $\delta R_{\rm s}$
\begin{equation}
\psi^{\rm opt}=g_1+g_2\sqrt{-\delta R_{\rm s}}+g_3 \delta R_{\rm s},   
\label{large-sigma-re-psi}
\end{equation}
where the coefficients $g_1$, $g_2$, and $g_3$ depend on $\tilde{P}^*$, $\varepsilon_{\rm C}$, $\eta_{\rm C}$, and $\Sigma_{\rm e}$ in a complicated way. Interestingly, for $\tilde{P}^*\to\infty$, $g_2$ vanishes and the increase in COP of the CAR becomes linear. As discussed at the end of the next section, for diverging $\tilde{P}^*$, the heat engine works at Carnot efficiency and the behavior of the MEGP of the CAR is solely determined by that of the refrigerator~\cite{holubec2020maximum}.

\begin{figure}[t]
\centering
\includegraphics[trim={0cm 0.5cm 0cm 0}, width=0.98\columnwidth]{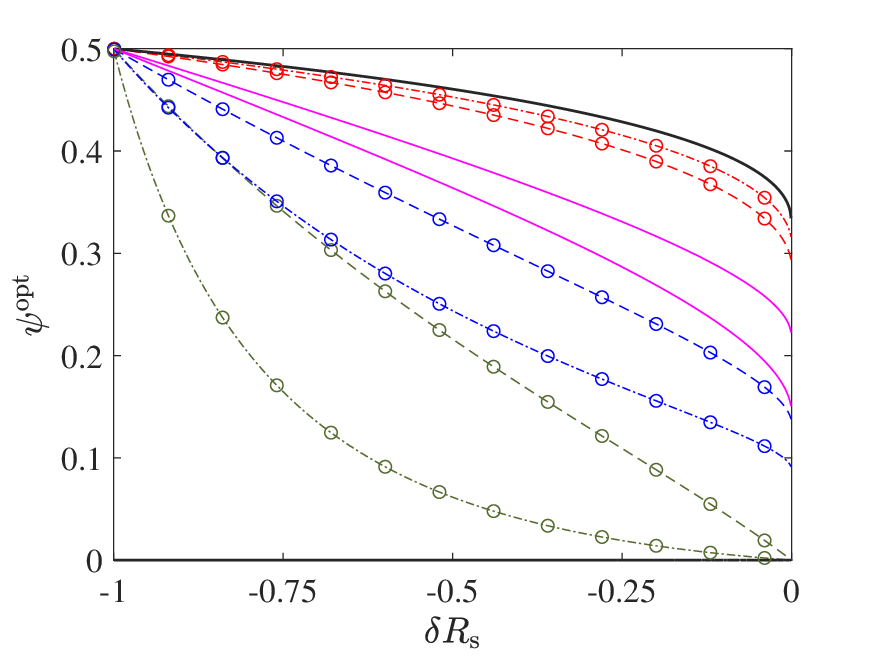}
\caption{The MEGP of the CAR~\eqref{final-COP-abso-refri} as a function of the loss in cooling power $\delta R_{\rm s}$ for three values 0, 1, and $\infty$ of the reduced maximum power of the engine $\tilde{P}^*$.
The reduced power $\tilde{P}^*$ increases from the uppermost
dashed line to the lowermost one with $\Sigma_{\rm r}=\Sigma_{\rm e}=1$. The dot-dashed lines of the same color as the dashed ones correspond to the same $\tilde{P}^*$ and $\Sigma_{\rm r}=\Sigma_{\rm e}=10$.
The pink solid lines depict the upper bound on MEGP~\eqref{ultimate-bounds} for fixed $\tilde{P}^*$ obtained for $\Sigma_{\rm r}=0$ and $\Sigma_{\rm e}\to\infty$. For the top one we took $\tilde{P}^*=1$. The bottom one corresponds to arbitrary $\tilde{P}^*\ge 1/\varepsilon_+^*$. The black solid lines represent the ultimate lower ($\Sigma_{\rm r}\to\infty$, $\Sigma_{\rm e}=0$, and arbitrary $\tilde{P}^*$) and upper ($\Sigma_{\rm r}=0$, $\Sigma_{\rm e}\to\infty$, and $\tilde{P}^*=0$) bounds on MEGP \eqref{extremely-ultimate-bounds}
(note that the lower bound coincides with the horizontal axis). MEGP for the CAR obtained using brute-force numerical optimisation of its COP (circles) perfectly agree with the curves calculated using the analytical formula~\eqref{final-COP-abso-refri} (lines). Other parameters taken: $\varepsilon_{\rm C}=1$ and $\eta_{\rm C}=1/2$.}
\label{fig:COP}
\end{figure}

\begin{figure*}[t]
\centering
\includegraphics[trim={4cm 1cm 3cm 4}, width=0.95\textwidth]{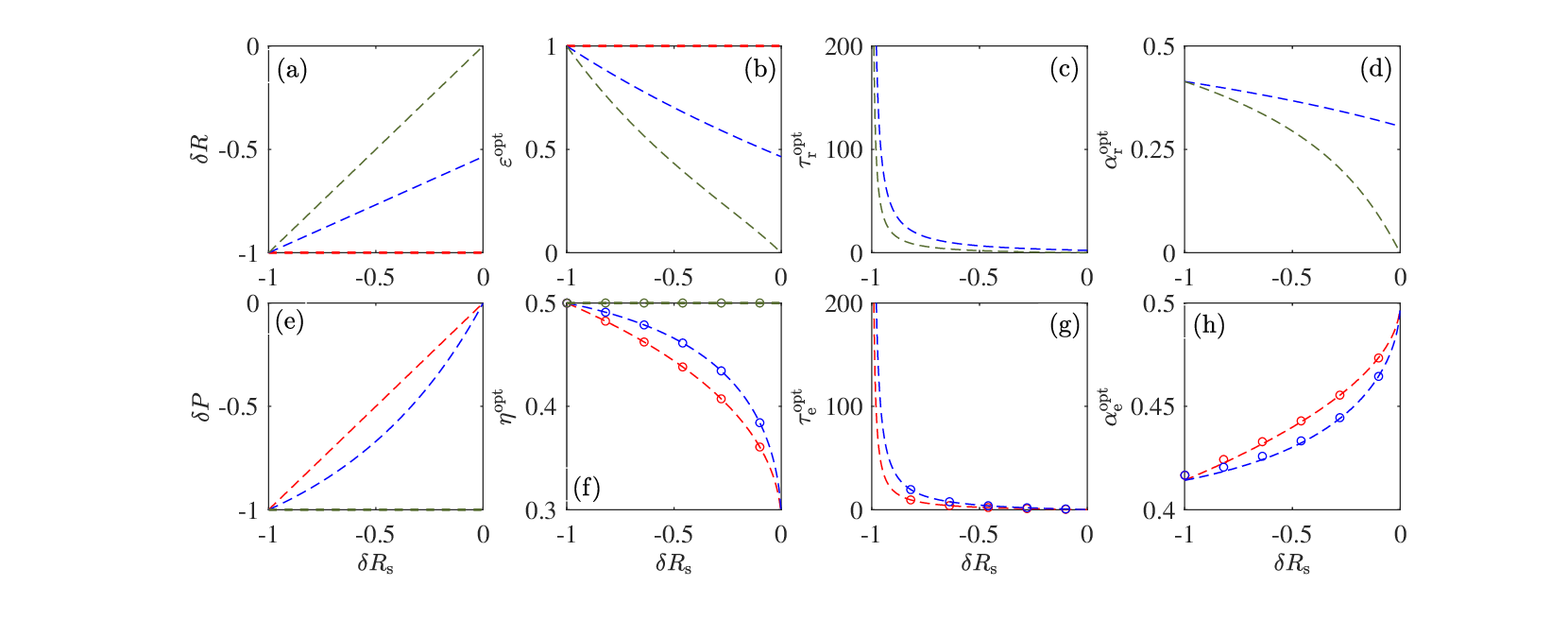}
\caption{Top: Characteristics of the internal refrigerator corresponding to the MEGP for the CAR depicted in Fig.~\ref{fig:COP} for $\Sigma_{\rm r}=\Sigma_{\rm e}=1$ as functions of the loss in output power of the CAR, $\delta R_{\rm s}$. (a) The loss in output power, $\delta R$, (b) MEGP, $\varepsilon^{\rm opt}$, (c) the optimal dimensionless cycle duration, $\tau_{\rm r}^{\rm opt}$, defined in Eq.~\eqref{tau-define}, and (d) the optimal relative duration of the hot isotherm, $\alpha_{\rm r}^{\rm opt}$, defined below Eq.~\eqref{tau-define}. Bottom: The corresponding characteristics of the internal heat engine. (e) The loss in output power, $\delta P$, (f) MEGP, $\eta^{\rm opt}$, (g) the optimal dimensionless cycle duration, $\tau_{\rm e}^{\rm opt}$ defined in Eq.~\eqref{tau-he-appxa}, and (h) the optimal relative duration of the hot isotherm, $\alpha_{\rm e}^{\rm opt}$, defined above Eq.~\eqref{appa-power}. Colors of the individual lines,
marking the used value of the reduced maximum power, $\tilde{P}^*$, are the same as in Fig.~\ref{fig:COP}  (red, blue, green corresponds to reduced powers 0, 1, and $\infty$, respectively). The vanishing reduced power corresponds to the Carnot COP of the refrigerator, $\varepsilon^{\rm opt} = \varepsilon_{\rm C}$, where
$\tau_{\rm r}^{\rm opt}$ diverges, and arbitrary $\alpha_{\rm r}^{\rm opt}$. Panels (c) and (d) thus show no red lines. Similarly, diverging $\tilde{P}^*$ corresponds to the Carnot efficiency of the engine, $\eta^{\rm opt}=\eta_{\rm C}$, and thus we show no green lines in panels (g) and (h). Even though the depicted parameters for the heat engines were obtained using the approximation \eqref{approximate-for-a}, they are almost indistinguishable from exact numerical results (circles). Slight deviations can be observed for $\alpha_{\rm e}^{\rm opt}$ only.}
\label{fig:Corresponding-parameters}
\end{figure*}

%%%%%%%%%%%%%%%%%%%%%%%%%%%%%%%%%%%%%%%%%%%%%%%%
\subsection{MEGP for arbitrary parameters}
\label{sec-a-effect}
%%%%%%%%%%%%%%%%%%%%%%%%%%%%%%%%%%%%%%%%%%%%%%%%

Outside the limiting parameter regimes discussed above, the \textcolor{black}{full analytical} expression~\eqref{final-COP-abso-refri} for the MEGP is too cumbersome to get an \textcolor{black}{immediate} insight into the behavior of $\psi^{\rm opt}$. Therefore, in this section, we investigate its dependence on the model parameters graphically.

In Fig. \ref{fig:lambda}, we show the reduced maximum power of the CAR, $\tilde{R}_{\rm s}^*$, as a function of the reduced power of the internal heat engine, $\tilde{P}^*$. The larger the available input power of the refrigerator (provided by the engine) the larger the corresponding maximum cooling power until the latter reaches its maximum, $\tilde{R}_{\rm s}^*=1$, where the whole cooling potential of the internal refrigerator is utilized. The minimum value of the reduced power, $\tilde{P}^* = 1/\varepsilon^{\rm opt} (0)$, allowing for $\tilde{R}_{\rm s}^*=1$ follows from Eq.~\eqref{rsx-rx}. Here, $\varepsilon^{\rm opt} (0)=\varepsilon^*_\pm$ denotes the MEGP for the refrigerator defined in Eqs. \eqref{final-COP-} and \eqref{final-COP+}. Hence, $\tilde{R}_{\rm s}^*=1$ for finite value ($\tilde{P}^*=1/\varepsilon_+^*$) of the reduced power only if $\Sigma_{\rm r}=0$, i.e., when the dissipation during the hot isotherm of the refrigeration cycle becomes negligible compared to that during the cold one. For $\Sigma_{\rm r}>0$, the whole cooling potential of the internal refrigerator can be utilized only for infinite values of the reduced power, $\tilde{P}^*=1/\varepsilon_-^*\to \infty$. This is caused by the discontinuity in the ability of the refrigerator working at maximum power conditions to utilize the energy provided by the engine, $\varepsilon^{\rm opt}(0)$, which is positive for $\Sigma_{\rm r} = 0$ and vanishes for $\Sigma_{\rm r} > 0$~\cite{holubec2020maximum}. Figure~\ref{fig:lambda} also shows that, for fixed $\tilde{P}^*$ and $\varepsilon_{\rm C}$,
$\tilde{R}_{\rm s}^*$ decreases as the amount of energy dissipated during the hot isotherm of the refrigeration cycle increases (larger $\Sigma_{\rm r}$).

In Fig.~\ref{fig:COP}, we plot the MEGP for CARs~\eqref{final-COP-abso-refri} as a function of $\delta R_{\rm s}$ for different values of $\tilde{P}^*$, $\Sigma_{\rm e}$, and $\Sigma_{\rm r}$.
\textcolor{black}{The upper bounds~\eqref{ultimate-bounds} for MEGP for fixed reduced power $\tilde{P}^*$ are depicted for $\tilde{P}^*=1$ (upper pink line) and $\tilde{P}^*\ge 1/\varepsilon_+^*$ (lower pink line). They indeed bound the MEGP obtained for arbitrary values of ratios of irreversibility parameters $\Sigma_{\rm e}$ and $\Sigma_{\rm r}$, and values of $\tilde{P}^*$ smaller than those chosen to plot the individual curves.} The ultimate upper bound on MEGP~\eqref{extremely-ultimate-bounds} is depicted by the uppermost black solid line. According to the figure, the MEGP $\psi^{\rm opt}$ exhibits a fast nonlinear increase with decreasing power near $\delta R_{\rm s}\to 0$ unless $\tilde{P}^*\to\infty$. Only then this increase is linear, in agreement with our discussion below Eq.~\eqref{large-sigma-re-psi}.
In order to check our analytical results, we also calculated the MEGP for simultaneous LD CARs by a direct brute-force numerical optimization of COP~\eqref{efficiency}. The figure shows that the obtained numerical results (symbols) perfectly overlap with our analytical predictions (lines).

In Fig.~\ref{fig:Corresponding-parameters}, we show the characteristics of the heat engine and refrigerator corresponding to the MEGP of the CAR with $\Sigma_{\rm r}=\Sigma_{\rm e}=1$, depicted in Fig.~\ref{fig:COP}.
For $\tilde{P}^*\to 0$, Eqs.~\eqref{third-relation}-\eqref{rsx-rx} imply that $\delta R=-1$,  $\varepsilon^{\rm opt}=\varepsilon_{\rm C}$, and $\delta P=\delta R_{\rm s}$.  Similarly, for $\tilde{P}^*\to \infty$ it follows that $\delta R=\delta R_{\rm s}$, $\delta P=-1$, and $\eta^{\rm opt}=\eta_{\rm C}$.  When the refrigerator works at the Carnot COP $\varepsilon_{\rm C}$, the dimensionless refrigeration cycle duration $\tau_{\rm r}^{\rm opt}$ diverges and we have $N_{\rm e}/N_{\rm r}\to\infty$, i.e. within one full CAR cycle, there is infinitely more engine cycles than refrigeration cycles. An opposite situation occurs when the engine works at Carnot efficiency.

%%%%%%%%%%%%%%%%%%%%%%%%%%%%%%%%%%%%%%%%%%%%%%%%%%%%%%%%%%%%%%%%%%%%%%%%%%%%%%%%%%%%%%%%%%%%%%%%
\section{Conclusion and Outlook}
\label{sec-conclusion}
%%%%%%%%%%%%%%%%%%%%%%%%%%%%%%%%%%%%%%%%%%%%%%%%%%%%%%%%%%%%%%%%%%%%%%%%%%%%%%%%%%%%%%%%%%%%%%%%

We have shown that the maximum efficiency at given cooling power (MEGP) for an absorption refrigerator composed of simultaneously operating Carnot-type heat engine and Carnot-type refrigerator (CAR) follows from the MEGPs for the internal heat engine and refrigerator. We have applied these general findings to low-dissipation (LD) simultaneous CARs, where the internal devices work in the LD regime and the corresponding expressions for MEGPs are known \cite{holubec2016maximum, holubec2020maximum}. We have used the resulting cumbersome analytical formula for the MEGP for derivation of concise expressions for upper and lower bounds on the MEGP for the LD CARs. We have also investigated behavior of the MEGP close to the maximum power. Unless the ratio of maximum powers of the internal engine and the refrigerator diverges, a slight decrease in power of the LD CAR leads to a fast nonlinear increase in the MEGP generically observed in heat engines~\cite{holubec2016maximum}. Otherwise, the increase in the MEGP is linear as can be observed in LD refrigerators~\cite{holubec2020maximum}.

In the LD approximation, the detailed dynamics of the system in question determines the so-called irreversibility parameters. The MEGP for simultaneous LD CARs, derived in this paper, is as function of power measured in units of the maximum power, which depends on the irreversibility parameters. Using a specific dynamical model, the maximum power can be further optimised with respect to theses parameters allowing to derive expressions for maximum power at fixed maximum efficiency. For LD heat engines and refrigerators, such an optimisation was performed in Refs.~\cite{PhysRevLett.124.110606,abiuso2020geometric} using the geometrical approach to thermodynamics generically valid close to equilibrium. While the dependence of maximum power on irreversibility parameters in these two settings is obvious, the situation in LD CARs is slightly different since their maximum power is controlled by both the maximum power of the internal refrigerator and that of the internal heat engine. Equations~\eqref{six-relation} and~\eqref{determine-allowed-Rs} suggest that the power of the CAR attains its maximum if one maximizes COP of the internal refrigerator, its maximum power, and also the maximum power of the heat engine. However, detailed investigations in this direction will be a subject of our future work.
 
The presented LD model is constructed in an idealized fashion and the resulting MEGP can serve as a (loose) upper bound for real-world absorption refrigerators. Such bounds are thus nowadays available for heat engines~\cite{holubec2016maximum}, refrigerators~\cite{holubec2020maximum}, and absorption refrigerators. It remains to derive them for heat pumps, which will also be a subject of our future work. For a numerical study of the MEGP for absorption heat pumps, we refer to Ref.~\cite{guo2020equivalent}. 
Furthermore, it would be interesting to investigate MEGPs for LD systems in context of the stability analysis described in Refs.~\cite{PhysRevLett.124.050603,PhysRevE.96.042128,PhysRevE.98.032142,PhysRevE.100.062128}.

\textcolor{black}{Originally, the finite-time performance of heat engines has been studied using the endoreversible model~\cite{curzon1975efficiency}. While efficiencies at maximum power for the endoreversible and LD models are described by similar expressions~\cite{Tu_2012}, \textcolor{black}{to the best of our knowledge,} no results for MEGP for endoreversible models are known. As a future research project, it would be also interesting to investigate to what extent the apparent equivalence between the two models holds concerning the MEGP.}

%%%%%%%%%%%%%%%%%%%%%%%%%%%%%%%%%%%%%%%%%%%%%%%%
\begin{acknowledgments}
We thank Paolo Abiuso and Mart\'i Perarnau-Llobet for bringing our attention to their work on optimization of power in slowly driven systems.
ZY also thanks Professor Jincan Chen of Xiamen University for instructive discussions of three-heat-source refrigerators and is grateful for the sponsorship of China Scholarship Council (CSC) under Grant No. 201906310136. VH gratefully acknowledges support by the Humboldt foundation and by the Czech Science Foundation (project No. 20-02955J).
\end{acknowledgments}
%%%%%%%%%%%%%%%%%%%%%%%%%%%%%%%%%%%%%%%%%%%%%%%%

\appendix

%%%%%%%%%%%%%%%%%%%%%%%%%%%%%%%%%%%%%%%%%%%%%%%%%%%%%%%%%%%%%%%%%%%%%%%%%%%%%%%%%%%%%%%%%%%%%%%%
\section{MEGP for alternating CARs}
\label{appx:method-comparison}
%%%%%%%%%%%%%%%%%%%%%%%%%%%%%%%%%%%%%%%%%%%%%%%%%%%%%%%%%%%%%%%%%%%%%%%%%%%%%%%%%%%%%%%%%%%%%%%%

For alternating CARs \cite{guo2019thermally}, the internal heat engine and refrigerator do not operate simultaneously. The duration of one cycle of the alternating CAR is thus given by the sum $t_{\rm e}+t_{\rm r}$ of the durations of the engine and refrigerator. According to the first law of thermodynamics, the output work of the heat engine per cycle equals to the input work of the refrigerator, i.e., $W_{\rm e}=W_{\rm r}$.  The power of the heat engine and the cooling power of the refrigerator then read
\begin{eqnarray}
\mathcal{P}&=&\frac{W_{\rm e}}{t_{\rm e}+t_{\rm r}}=\frac{P}{1+\Lambda},
\label{appxc-tilde-heat-power}
\\
\mathcal{R}&=&\frac{Q_{\rm c}}{t_{\rm e}+t_{\rm r}}=\frac{R}{1+1/\Lambda},
\label{appxc-tilde-cooling-power}
\end{eqnarray}
where $\Lambda\equiv t_{\rm r}/t_{\rm e}$ measures the ratio of durations of the two internal cycles. The first law in the form of Eq.~\eqref{second-relation} implies that these two powers are interconnected through the COP $\varepsilon$ of the refrigerator, 
%Since the parameter $t_{\rm e}$ is not involved in the COP of the internal refrigerator~\eqref{ini-appa-cooling-power-refrigerator} , $\varepsilon$, Eq. \eqref{second-relation} is rewritten as
\begin{equation}
\mathcal{P}=\frac{\mathcal{R}}{\varepsilon(R)}. 
\label{appxc-tilde-first-law}
\end{equation}
%Similarly, the parameter $t_{\rm r}$ is not involved in efficiency \eqref{appa-power-engine}, $\eta$.
Using Eqs. \eqref{appxc-tilde-cooling-power} and \eqref{appxc-tilde-first-law}, the COP~\eqref{efficiency} can be rewritten as
\begin{equation}
\begin{split}
\psi(\mathcal{R})&=\varepsilon(R)\eta(P)\\
&=\varepsilon(\mathcal{R}, \Lambda)\eta(\mathcal{P}, \Lambda)\\
&=\varepsilon(\mathcal{R}, \Lambda)\eta\left(\frac{\mathcal{R}}{\varepsilon(\mathcal{R}, \Lambda)}, \Lambda\right),
\label{appxc-tilde-efficiency}
\end{split}
\end{equation}
where the notation $\varepsilon(R) = \varepsilon[\mathcal{R}(1 + 1/\Lambda)] \equiv \varepsilon(\mathcal{R}, \Lambda)$ highlights that both the efficiencies now explicitly depend on the ratio of the durations of the internal cycles, $\Lambda$, through the definitions \eqref{appxc-tilde-heat-power} and \eqref{appxc-tilde-cooling-power} of $\mathcal{R}$ and $\mathcal{P}$. Consequently, in the optimisation of COP~\eqref{appxc-tilde-efficiency} with respect to the durations of the refrigeration and engine cycles, the engine and refrigeration efficiencies can not be optimised independently as it was done in Eq.~\eqref{opt-opt-relation-engine-refrigerator}. 
 The optimisation of COP for alternating CARs is thus more complicated than that for simultaneous CARs and the knowledge of MEGPs for the internal engine and refrigerator might not be sufficient for determination of MEGP for alternating CARs.

%%%%%%%%%%%%%%%%%%%%%%%%%%%%%%%%%%%%%%%%%%%%%%%%%%%%%%%%%%%%%%%%%%%%%%%%%%%%%%%%%%%%%%%%%%%%%%%%
\section{MEGP for LD heat engines}
\label{appx:opt-engine}
%%%%%%%%%%%%%%%%%%%%%%%%%%%%%%%%%%%%%%%%%%%%%%%%%%%%%%%%%%%%%%%%%%%%%%%%%%%%%%%%%%%%%%%%%%%%%%%%

In this appendix, we derive the expression $\eta^{\rm opt}$ for MEGP for LD heat engines. The derivation is slightly different from that used in Ref.~\cite{holubec2016maximum}.

Introducing the relative duration of the hot isotherm, $\alpha_{\rm e}=t_{\rm h}/t_{\rm e}$, in Eqs. \eqref{qh} and \eqref{qoh}, the power output and efficiency of the LD heat engine can be expressed as 
\begin{eqnarray}
P&=&\frac{W_{\rm e}}{t_{\rm e}}=\frac{(T_{\rm h}-T_{\rm m})\Delta S_{\rm e}}{t_{\rm e}}-\frac{\alpha\Sigma_{\rm me}+(1-\alpha)\Sigma_{\rm h}}{\alpha(1-\alpha)t_{\rm e}^2}, 
\label{appa-power}\\
\eta&=&\frac{W_{\rm e}}{Q_{\rm h}}=\frac{\eta_{\rm C}}{1+T_{\rm m}\Delta S_{\rm tot, e}/(Pt_{\rm e})},
\label{appa-power-engine}
\end{eqnarray}
where
\begin{equation}
\textcolor{black}{
\Delta S_{\rm tot, e}=-\frac{Q_{\rm h}}{T_{\rm h}}+\frac{Q_{\rm me}}{T_{\rm m}}=\frac{\Sigma_{\rm h}}{t_{\rm h} T_{\rm h}}+\frac{\Sigma_{\rm me}}{t_{\rm me}T_{\rm m}}\ge 0} 
\label{deltaS-he}
\end{equation}
is the total entropy production per engine cycle.

Maximizing the power \eqref{appa-power} with respect to $\alpha$ and $t_{\rm e}$ yields \cite{schmiedl2007efficiency}
\begin{eqnarray}
\alpha_{\rm e}^*&=&\frac{\sqrt{\Sigma_{\rm e}}}{1+\sqrt{\Sigma_{\rm e}}},
\label{appa-alpha-star}\\
t_{\rm e}^*&=&\frac{2\left(\sqrt{\Sigma_{\rm h}}+\sqrt{\Sigma_{\rm me}}\right)^2}{T_{\rm h}\eta_{\rm C}\Delta S_{\rm e}},  
\label{appa-time-star}\\
P^*&=&\frac{1}{4}\left(\frac{T_{\rm h}\eta_{\rm C}\Delta S_{\rm e}}{\sqrt{\Sigma_{\rm h}}+\sqrt{\Sigma_{\rm me}}}\right)^2, 
\label{appa-power-engine-star}\\
\eta^*&=&\frac{\eta_{\rm C}\left(1+\sqrt{\Sigma_{\rm e}}\right)}{2+\sqrt{\Sigma_{\rm e}}(2-\eta_{\rm C})},   
\label{app-EMP}
\end{eqnarray}
where $\Sigma_{\rm e}=\Sigma_{\rm h}/\Sigma_{\rm me}$ is the so-called irreversibility ratio and $\eta_{\rm C}=1-T_{\rm m}/T_{\rm h}$ denotes Carnot efficiency.
Now, we use Eqs. \eqref{appa-alpha-star} and \eqref{appa-time-star} to define the coordinate transformation
\begin{eqnarray}
\tau_{\rm e}&=&\frac{t_{\rm e}}{t_{\rm e}^*}-1~\in[-1, \infty],   
\label{tau-he-appxa}\\
a&=&\frac{\alpha_{\rm e}}{\alpha_{\rm e}^*}-1~\in\left[-1, \frac{1}{\alpha_{\rm e}^*}-1\right],
\label{a-he-appxa}
\end{eqnarray}
which reduces the number of variables in the problem \cite{holubec2016maximum}. The point of maximum power \eqref{appa-power-engine-star} corresponds to $\delta P=0$ \eqref{coordinate} and $\tau_{\rm e}=a=0$.
The (relative) loss in power \eqref{coordinate} and efficiency \eqref{appa-power-engine} in these new coordinates read
\begin{equation}
\delta P=\frac{a^2\sqrt{\Sigma_{\rm e}}}{(1+a)(a\sqrt{\Sigma_{\rm e}}-1)(1+\tau_{\rm e})^2}-\left(\frac{\tau_{\rm e}}{1+\tau_{\rm e}}\right)^2,
\label{appa-relative-power-final-engine}
\end{equation}
\begin{equation}
\begin{split}
\eta&=\frac{(1+\sqrt{\Sigma_{\rm e}})\eta_{\rm C}}{a\sqrt{\Sigma_{\rm e}}-1}\\
&\times\frac{2a^2\sqrt{\Sigma_{\rm e}}(1+\tau_{\rm e})+(a\sqrt{\Sigma_{\rm e}}-a-1)(1+2\tau_{\rm e})}{2(1+a)(1+\sqrt{\Sigma_{\rm e}})(1+\tau_{\rm e})-\eta_{\rm C}\sqrt{\Sigma_{\rm e}}}.
\end{split}
\label{appa-relative-effi-final-engine}
\end{equation}
Solving Eq.~\eqref{appa-relative-power-final-engine} with respect to the dimensionless cycle duration $\tau_{\rm e}$, we find two roots
\begin{equation}
\tau_{\rm e}=\frac{-\delta P}{1+\delta P}\pm\frac{\sqrt{\delta P(1+a-a\sqrt{\Sigma_{\rm e}})+a^2\sqrt{\Sigma_{\rm e}}}}{(1+\delta P)\sqrt{(1+a)(a\sqrt{\Sigma_{\rm e}}-1)}}.
\label{root-for-tau-engine}
\end{equation}
%which mark the different combinations of coordinates $\tau_{\rm e}$ and $a$ yielding the same value of power. 
Since longer cycles
in general allow for larger efficiencies, and thus we take the root with the positive sign. Substituting it %the desired root of $\tau_{\rm e}$ \eqref{root-for-tau-engine} 
into Eq.~\eqref{appa-relative-effi-final-engine}, evaluating the condition $\left.\partial\eta/\partial a\right|_{a=a^{\rm opt}} = 0$ for maximum efficiency, and expanding it up to the fourth order in $a$, we find
\begin{equation}
\left.\frac{\partial\eta}{\partial a}\right|_{a=a^{\rm opt}}=\sum_{\rm n=0}^{4}b_na^n+\mathcal{O}(a^5)=0,   
\label{approximate-for-a}
\end{equation}
where the coefficients $b_n$ are complicated functions of $\delta P$, $\Sigma_{\rm e}$, and $\eta_{\rm C}$.
The equation~\eqref{approximate-for-a} for the optimal value $a^{\rm opt}$ of the parameter $a$ can be solved exactly~\cite{abramowitz1964}. The corresponding optimal value of $\tau^{\rm opt}_{\rm e}$ follows by substituting the resulting $a^{\rm opt}$ for $a$ in Eq.~\eqref{root-for-tau-engine}.

Substituting the obtained expressions for $a^{\rm opt}$ and $\tau^{\rm opt}_{\rm e}$ for $a$ and $\tau$ into Eq.~\eqref{appa-relative-effi-final-engine}, we obtain \textcolor{black}{a lengthy but manageable, e.g. by using software for symbolic manipulation such as Mathematica, formula} for the MEGP for LD heat engines
\begin{equation}
\eta^{\rm opt}=\eta^{\rm opt}(\delta P, \Sigma_{\rm e}, \eta_{\rm C}).
\label{appa-eta-opt-del-power-f3}
\end{equation}
Even though this results was obtained using the approximation~\eqref{approximate-for-a}, we have tested that the resulting approximate MEGP \eqref{appa-eta-opt-del-power-f3} and the exact MEGP obtained numerically are indistinguishable within the numerical precision \textcolor{black}{(the measured absolute error is on the order of 10$^{-7}$)}. 
Furthermore, the expression \eqref{appa-eta-opt-del-power-f3} yields exact lower ($\Sigma_{\rm e}=0$) and upper ($\Sigma_{\rm e}\to \infty$) bounds on the MEGP of LD heat engines~\cite{holubec2016maximum}
\begin{equation}
\frac{\eta_{\rm C}}{2}\left(1+\sqrt{-\delta P}\right)\le\eta^{\rm opt}\le\frac{\eta_{\rm C}(1+\sqrt{-\delta P})}{2-\eta_{\rm C}(1-\sqrt{-\delta P})}.
\label{bounds-heat-engine}
\end{equation}

%%%%%%%%%%%%%%%%%%%%%%%%%%%%%%%%%%%%%%%%%%%%%%%%
%%%%%%%%%%%%%%%%%%%%%%%%%%%%%%%%%%%%%%%%%%%%%%%%
\section{MEGP for LD refrigerators}
\label{appx:opt-refrigerator}
%%%%%%%%%%%%%%%%%%%%%%%%%%%%%%%%%%%%%%%%%%%%%%%%
%%%%%%%%%%%%%%%%%%%%%%%%%%%%%%%%%%%%%%%%%%%%%%%%

In this appendix, we review the derivation of the expression $\varepsilon^{\rm opt}$ for MEGP for LD refrigerators given in Ref.~\cite{holubec2020maximum}.

The COP of the refrigerator is given by
\begin{equation}
\varepsilon=\frac{Q_{\rm c}}{W_{\rm r}}=\frac{\varepsilon_{\rm C}}{1+\varepsilon_{\rm C}T_{\rm m}\Delta S_{\rm tot, r}/(Rt_{\rm r})}, 
\label{ini-appa-cooling-power-refrigerator}
\end{equation}
where
\begin{equation}
\textcolor{black}{
\Delta S_{\rm tot, r}=-\frac{Q_{\rm c}}{T_{\rm c}}+\frac{Q_{\rm mr}}{T_{\rm m}}=\frac{\Sigma_{\rm c}}{t_{\rm c} T_{\rm c}}+\frac{\Sigma_{\rm mr}}{t_{\rm mr}T_{\rm m}}\ge 0}
\label{deltaS-re}
\end{equation}
is the total entropy production per refrigeration cycle.
Substituting Eq.~\eqref{qc} into Eq.~\eqref{power-tp} and maximizing the resulting expression with respect to $t_{\rm mr}$ and $t_{\rm c}$ gives \cite{holubec2020maximum, hernandez2015time}
\begin{eqnarray}
t_{\rm c}^*&=&t_{\rm r}^*=\frac{2\Sigma_{\rm c}}{T_{\rm c}\Delta S_{\rm r}}, 
\label{appb-time-star}\\
R^*&=&\frac{(T_{\rm c}\Delta S_{\rm r})^2}{4\Sigma_{\rm c}}.
\label{appb-power-star}
\end{eqnarray}
At maximum power conditions, the duration of the cold isotherm $t_{\rm c}^*$ thus equals to the duration of the whole cycle $t_{\rm r}^*$, which should be understood in the sense that the hot isotherm is infinitely faster than the cold one. The corresponding COP of the internal refrigerator at maximum power, $\varepsilon^*$, reads
\begin{eqnarray}
\varepsilon_-^ * &=& 0 \quad \quad \quad \,\,\, {\rm for} \quad \Sigma_{\rm r}>0
\label{final-COP-},\\
\varepsilon_+^ * &=& \frac{\varepsilon_{ \rm C}}{2+\varepsilon_{\rm C}} \quad {\rm for} \quad \Sigma_{\rm r}=0,
\label{final-COP+}
\end{eqnarray}
where $\Sigma_{\rm r}=\Sigma_{\rm mr}/\Sigma_{\rm c}$ is the so-called irreversibility ratio and $\varepsilon_{\rm C}=T_{\rm c}/(T_{\rm m}-T_{\rm c})$ denotes Carnot COP. The COP at maximum power $\varepsilon^*$ thus exhibits a discontinuity at $\Sigma_{\rm r}=0$.
Using Eq.~\eqref{appb-time-star}, we define the dimensionless cycle duration as
\begin{equation}
\tau_{\rm r}=\frac{t_{\rm r}}{t_{\rm r}^*}-1~\in[-1, \infty].  
\label{tau-define}
\end{equation}
Introducing further the relative duration of the hot isotherm $\alpha_{\rm r}=t_{\rm mr}/t_{\rm r}$, we find from Eqs.~\eqref{power-tp}, \eqref{coordinate} and \eqref{tau-define} that
\begin{equation}
\alpha_{\rm r}=1+\frac{1}{\left(1+\delta R\right)\tau_{\rm r}^2+2\delta R\tau_{\rm r}+\delta R-1}.
\label{alpha-deltaR-tau}
\end{equation}
Since $\alpha_{\rm r}$ by definition satisfies $0\le\alpha_{\rm r}\le 1$, the above formula makes sense only if
\begin{equation}
-\frac{\sqrt{-\delta R}}{{1+\sqrt { -\delta R}}} \le \tau_{\rm r}  \le \frac{\sqrt{-\delta R}}{{1-\sqrt { -\delta R}}}.
\label{tau-duration}
\end{equation}
The COP~\eqref{ini-appa-cooling-power-refrigerator} in these new variables reads
\begin{equation}
\varepsilon  = \frac{\tau_{\rm r} ^3 +A_{1,3}\tau_{\rm r} ^2 +A_{0,3}\tau_{\rm r}  + A_{0,1}}{- {\tau_{\rm r} ^3} + A_{1/{\varepsilon_+^ *},-3} \tau_{\rm r}^2 + B_{3,4,1}\tau_{\rm r} + B_{1,2,-1}},
\label{cop-deltap-equation}
\end{equation}
with $A_{k,l} = (k +l \delta R)/(1 + \delta R)$ and $B_{k,l, m}= [-k{\left( \delta R \right)}^2+ \left( l/\varepsilon _{\rm C} + 1+\Sigma_{\rm r} \right)\delta R+ m \Sigma_{\rm r}]/ \left(1 + \delta R \right)^2$.
The maximum of COP~\eqref{cop-deltap-equation} can be determined by the condition $\partial \varepsilon /\partial \tau_{\rm r}|_{\tau_{\rm r}=\tau_{\rm r}^{\rm opt}}  = 0$, which explicitly reads
\begin{multline}
(\tau_{\rm r}^{\rm opt}) ^4 + \tilde{A} (\tau_{\rm r}^{\rm opt})  ^3 + \tilde{B}_{6+3\tilde{\Sigma}_{\rm r},2+2\tilde{\Sigma}_{\rm r},-\tilde{\Sigma}_{\rm r}} (\tau_{\rm r}^{\rm opt})  ^2\\ 
+ \tilde{B}_{4+3\tilde{\Sigma}_{\rm r},-2\tilde{\Sigma}_{\rm r},-\tilde{\Sigma}_{\rm r}}\tau_{\rm r}^{\rm opt}  
+ \tilde{B}_{1+\tilde{\Sigma}_{\rm r},-2\tilde{\Sigma}_{\rm r},0} = 0.
\label{max-cop-derive}
\end{multline}
Above, the coefficients $\tilde{A} = [(4+\tilde{\Sigma}_{\rm r})\delta R + \tilde{\Sigma}_{\rm r}]/(1+\delta R)$ and $\tilde{B}_{k,l,m} = (k\delta R^2 + l \delta R + m)/(1+\delta R)^2$ depend on $\Sigma_{\rm r}$ and $\varepsilon _{\rm C}$ only through the combination $\tilde \Sigma_{\rm r} = \Sigma_{\rm r} /\left({\frac{1}{{{\varepsilon _{\rm C}}}} + 1}\right)$.

The quartic equation~\eqref{max-cop-derive} has four roots and can be analytically solved~\cite{abramowitz1964,holubec2020maximum}. The optimal dimensionless cycle duration $\tau_{\rm r}^{\rm opt}=\tau_{\rm r}^{\rm opt}(\delta R, \Sigma_{\rm r}, \varepsilon_{\rm C})$ is determined by the only physically reasonable root, located in the interval \eqref{tau-duration}. Substituting it for $\tau$ in Eq.~\eqref{cop-deltap-equation},
we obtain \textcolor{black}{a lengthy but manageable, e.g. by using software for symbolic manipulation such as Mathematica,} exact expression for $\varepsilon^{\rm opt}$,
\begin{equation}
\varepsilon^{\rm opt}=\varepsilon^{\rm opt}(\delta R, \Sigma_{\rm r}, \varepsilon_{\rm C}). 
\label{eta-opt-appb-m}
\end{equation}
It turns out to be bounded by the inequalities
\begin{equation}
0\le\varepsilon^{\rm opt} \le \frac{\varepsilon_{\rm C}(1+\sqrt{-\delta R})}{2+\varepsilon_{\rm C}(1-\sqrt{-\delta R})}\equiv\varepsilon_+^{\rm opt},
\label{eta-duration2}
\end{equation}
where the lower bound corresponds to $\Sigma_{\rm r}\to \infty$ and the upper bound to $\Sigma_{\rm r}=0$.

\bibliography{References}
\end{document}